\documentstyle[epsf,12pt]{article}
\def\photonatomright{\begin{picture}(3,1.5)(0,0)
                                \put(0,-0.75){\tencircw \symbol{2}}
                                \put(1.5,-0.75){\tencircw \symbol{1}}
                                \put(1.5,0.75){\tencircw \symbol{3}}
                                \put(3,0.75){\tencircw \symbol{0}}
                      \end{picture}
                     }

\def\photonright{\begin{picture}(30,1.5)(0,0)
                     \multiput(0,0)(3,0){10}{\photonatomright}
                  \end{picture}
                 }

\def\photonrighthalf{\begin{picture}(30,1.5)(0,0)
                     \multiput(0,0)(3,0){5}{\photonatomright}
                  \end{picture}
                 }

\def\fermionuphalf{\begin{picture}(1,15)(0,0)
                         \put(0,0){\vector(0,1){7.5}}
                         \put(0,7.5){\line(0,1){7.5}}
                   \end{picture}
                  }

\def\fermiondown{\begin{picture}(1,30)(0,-30)
                       \put(0,0){\vector(0,-1){15}}
                       \put(0,-15){\line(0,-1){15}}
                 \end{picture}
                }

\def\fermionull{\begin{picture}(30,15)(0,0)
                        \put(0,0){\vector(-2,1){15}}
                        \put(-15,7.5){\line(-2,1){15}}
                  \end{picture}
                 }
\def\fermionullhalf{\begin{picture}(15,7.5)(0,0)
                        \put(0,0){\vector(-2,1){7.5}}
                        \put(-7.5,3.75){\line(-2,1){7.5}}
                  \end{picture}
                 }
\def\fermionurr{\begin{picture}(30,15)(0,0)
                        \put(-30,-15){\vector(2,1){15}}
                        \put(-15,-7.5){\line(2,1){15}}
                  \end{picture}
                 }
\def\fermionurrhalf{\begin{picture}(15,7.5)(0,0)
                        \put(-15,-7.5){\vector(2,1){7.5}}
                        \put(-7.5,-3.75){\line(2,1){7.5}}
                  \end{picture}
                 }
\def\fermiondrr{\begin{picture}(30,15)(0,0)
                        \put(0,0){\vector(2,-1){15}}
                        \put(15,-7.5){\line(2,-1){15}}
                  \end{picture}
                 }
\def\fermiondrrhalf{\begin{picture}(15,7.5)(0,0)
                        \put(0,0){\vector(2,-1){7.5}}
                        \put(7.5,-3.75){\line(2,-1){7.5}}
                  \end{picture}
                 }
\def\fermiondll{\begin{picture}(30,15)(0,0)
                        \put(30,15){\vector(-2,-1){15}}
                        \put(15,7.5){\line(-2,-1){15}}
                  \end{picture}
                 }
\def\fermiondllhalf{\begin{picture}(15,7.5)(0,0)
                        \put(15,7.5){\vector(-2,-1){7.5}}
                        \put(7.5,3.75){\line(-2,-1){7.5}}
                  \end{picture}
                 }

\newenvironment{Feynman}[3]{\begin{center}
                            \setlength{\unitlength}{#3 mm}
                            \begin{picture}(#1)(#2)
                            \thicklines
                           }{\end{picture} \end{center}}
\newcommand{\nn}{\noindent}
\newcommand{\nll}{\nonumber \\}

\newcommand{\bq}{\begin{equation}}
\newcommand{\eq}{\end{equation}}
\newcommand{\ba}{\begin{eqnarray}}
\newcommand{\ea}{\end{eqnarray}}

\newcommand{\nobody}{\rule{0ex}{1ex}}

\begin{document}
\thispagestyle{empty}
\thispagestyle{empty}
\onecolumn
\date{}
\vspace{-1.4cm}
\begin{flushleft}
{DESY 94--185 \\}
{CERN--TH. 7478/94 \\}
{LMU 18/94 \\}
 October 1994
\end{flushleft}
\vspace{1.5cm}
\begin{center}
{\LARGE 
The process
$\;e^+ e^- \rightarrow l  \bar l \; q  \bar q $
\, at LEP and NLC
}
\vspace*{1.0cm}
\vfill
\nn
{\large
Dima Bardin$^{1,2 }$, $\;$ Arnd Leike$^{3\sharp}$
$\; $ and $\;$ Tord~Riemann$^4$}
\\
\vspace*{1.0cm}
\end{center}

\begin{itemize}
\item[$^1$]
Theoretical Physics Division, CERN, CH--1211 Geneva 23, Switzerland
\item[$^2$]
Theoretical Physics Laboratory, JINR,
ul. Joliot-Curie 6, 
RU--141980 Dubna,
\\ Moscow Region, Russia
\item[$^3$]
Lehrstuhl Prof. Fritzsch, Sektion Physik der
Ludwig-Maximilians-Universit\"at,                 \\
Theresienstr. 37, D--80333 M\"unchen, Germany
\item[$^4$]
DESY--Institut f\"ur Hochenergiephysik, Platanenallee 6, D--15738
Zeuthen, Germany
\end{itemize}
\vfill
\vspace*{1.0cm}

\centerline{\Large
Abstract}
\vspace*{.3cm}
\nn
The cross sections for the reaction
$e^+\;e^- \rightarrow l\;\bar l\;q\;\bar q $ and for similar four fermion
production
processes at LEP~1, LEP~2 and the NLC
are calculated.
Due to the extraordinary symmetry properties of the process,
very compact analytical formulae describe the
double differential distributions in the
invariant masses of the $l \bar l$ and $q \bar{q}$ pairs.
The total cross sections may be obtained with two numerical integrations.
\vspace*{1.0cm}
\vfill
\footnoterule
\nn
{\small
{\tt
{email:\ \
bardindy@cernvm.cern.ch, leike@cernvm.cern.ch, riemann@ifh.de\\
}
\nobody{\normalsize $^{\sharp}$} {\footnotesize
Supported by the German Federal Ministry for Research
and Technology under contract No.~05~GMU93P.}
} }
\pagebreak
\section{Introduction}
At LEP~1 energies, few events of
electron positron annihilation into four fermions, $e^+e^- \rightarrow
4f$,
have been observed~\cite{lep1zz} although $4f$ production
is extremely suppressed compared to fermion pair production.
The 4$f$ production is also interesting as a
background contribution to light Higgs searches~\cite{lep1zh}
with the associated $ZH$ production.
The strong kinematical suppressions are hoped to become overcompensated
by the rising number of $Z$ bosons being produced
at LEP~1.

Above the production thresholds for gauge boson pairs,
four fermion production is predicted by the Standard Model to
be one of the most frequent annihilation processes.
It will allow to study further details of
the gauge boson properties; and
the search for Higgs boson signals from
the Bjorken process will be one of the challenges of
LEP~2.

Several
Monte Carlo approaches to the complete description of 4$f$ production
have been developed~\cite{kleiss,kleiss2,booszz,pittau} while we
are performing a program of {\it semi-analytical} calculations
{}~\cite{wwqed,wwteup,zhteu}.
Although the treatment of even the simplest final states
is technically involved
we got encouraging results for off shell $W$ pair
production~\cite{wwqed,wwteup}
and the Bjorken process (off shell $ZH$ production)~\cite{zhteu}.
Some general remarks on $4f$ production and numerical comparisons may be found
in~\cite{pittau,wwteup,bglasg}.
The neutral current (NC) channels for $4f$ final states are classified in
table~1.
The four different event classes                          are discussed
in~\cite{wwteup}.
Higgs boson exchange has been excluded;
this is a very good
approximation if one doesn't perform a dedicated Higgs
search~\cite{zhteu}.
\begin{table}[b]
\vspace{-.25cm}
\label{tab2}
\begin{center}
 \begin{tabular}{|c|c|c|c|c|c|c|}
\hline
&
\raisebox{0.pt}[2.5ex][0.0ex]{${\bar d} d$}
&${\bar u} u$
&${\bar e} e$
&${\bar \mu} \mu$
&${\bar \nu}_{e} \nu_{e}$
&${\bar \nu}_{\mu} \nu_{\mu}$
\\
\hline
\raisebox{0.pt}[2.5ex][0.0ex]{${\bar d} d$}
 & {\tt 4$\cdot $16} & {\it 43} & {48}
             & {\bf 24} & 21 & {\bf 10} \\
\hline
\raisebox{0.pt}[2.5ex][0.0ex]
{${\bar s} s, {\bar b} b$} & {\bf 32} & {\it 43} & {48}
             & {\bf 24} & {21} & {\bf 10} \\
\hline
${\bar u} u$ & {\it 43} & {\tt 4$\cdot$16} & {48}
             & {\bf 24} & {21} & {\bf 10} \\
\hline
${\bar e} e$ &{48} &{48} & {\tt 4$\cdot$36} &{48}
& {\it 56} & {20}
\\
\hline
${\bar \mu} \mu$  & {\bf 24} & {\bf 24} & {48} & {\tt 4$\cdot$12}
                  & {19} & {\it 19}         \\
\hline
${\bar \tau} \tau$& {\bf 24} & {\bf 24} & {48} & {\bf 24}
                  & {19} & {\bf 10}         \\
\hline
${\bar \nu}_e \nu_{e}$  & {21} & {21} & {\it 56} & {19}
                  & {\tt 4$\cdot$9} & {12}                   \\
\hline
${\bar \nu}_{\mu} \nu_{\mu}$ & {\bf 10} & {\bf 10} & {20}
             & {\it 19} & {12} & {\tt 4$\cdot$3}  \\
\hline
${\bar \nu}_{\tau} \nu_{\tau}$ & {\bf 10} & {\bf 10} & {20}
             & {\bf 10} & {12} & {\bf 6}  \\
\hline
\end{tabular}
\vspace{.2cm} \\
\caption[]
{\it
Number of Feynman diagrams for
`NC' type final states.
}
\end{center}
\vspace{-.5cm}
\end{table}

In this article, we deal with the simplest event
class marked in {\bf boldface}.
It comprises final states which do
neither contain electrons (positrons), electron (anti)neutrinos, nor
identical fermions and cannot be produced by charged current interactions.

Our calculation covers the following {\em observable} final states:
\begin{itemize}
\item[(i)]
[$\mu \bar \mu, \tau \bar \tau$],
\item[(ii)]
[$l \bar l, b \bar b$],
[$l \bar l, c \bar c$],
[$l \bar l, (u \bar u + d \bar d + s \bar s + c \bar c + b \bar b)$],
\item[(iii)]
[$b \bar b, c \bar c$],
\end{itemize}
where $l= \mu, \tau$.
Other reactions of table~1, e.g. those with $\nu_{\tau}$ and $\nu_{\mu}$
pairs, may also be
calculated with the below formulae but are not observable due to the admixture
of other final states, here e.g. $\nu_{e}$ pairs. The latter belong to another,
much more complicated type of reaction.

In section~2, the calculation is shortly described.
Section~3 contains the main result: the analytical formulae for the invariant
fermion pair mass distributions.
In section~4 we discuss numerical results.

\section{Feynman diagrams, phase space, and cross section}
The  reaction
\ba
e^+e^- \rightarrow l {\bar l} + b {\bar b}
\label{lb}
\ea
is representative for the class of reactions considered here.
It is described
by 24 Feynman diagrams with six different topologies which we will call
{\tt crab} diagrams and {\tt deer} diagrams; see figures~1 and~2.
If there are two quark pairs in the final state, the {\tt deers} may also
contain gluon exchange.
This will be discussed later.

We parametrize the eightdimensional four particle phase space as follows:
%
\begin{eqnarray}
\label{domega}
d\Gamma
&=& \prod_{i=1}^4\frac{d^3p_i}{2p_{i}^0}
 \times  \delta^4(k_1+k_2-\sum_{i=1}^4 p_i)
\nonumber
\\
&=&~2\pi\frac{\sqrt{\lambda(s,s_1,s_2)}}{8s}
\frac{\sqrt{\lambda(s_1,m_1^2,m_2^2)}}{8s_1}
\frac{\sqrt{\lambda(s_2,m_3^2,m_4^2)}}{8s_2}
d s_1 d s_2 d \cos\theta d \Omega_1 d \Omega_2,
\end{eqnarray}
with the usual definition of the $\lambda$ function,
\ba
\lambda(a,b,c) &=& a^2+b^2+c^2-2ab-2ac-2bc,
\\
\lambda &\equiv& \lambda(s,s_1,s_2).
\label{lambda}
\ea

\begin{figure}[bhtp]
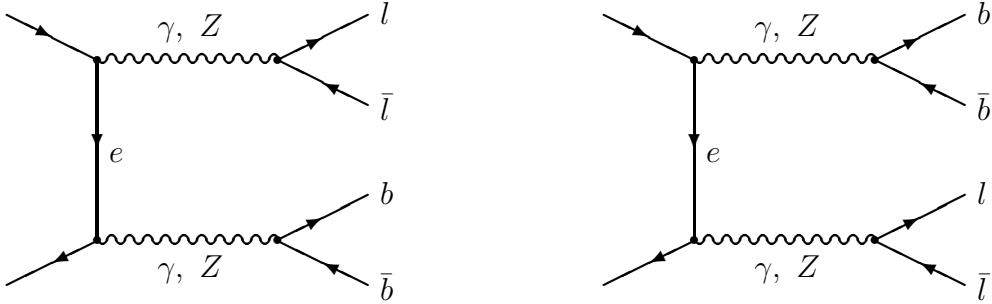

\begin{minipage}[tbh]{7.8cm} {
\begin{center}
\begin{Feynman}{75,60}{0,0}{0.8}
%
\put(15,7.5){\fermiondllhalf}
\put(15,52.5){\fermiondrrhalf}
\put(30,15){\fermiondown}
\put(30,45){\photonright}
\put(30,15){\photonright}
\put(30,45){\circle*{1.5}}
\put(30,15){\circle*{1.5}}
\put(60,15){\circle*{1.5}}
\put(60,45){\circle*{1.5}}
\put(75,7.5){\fermionullhalf}
\put(75,22.5){\fermionurrhalf}
\put(75,37.5){\fermionullhalf}
\put(75,52.5){\fermionurrhalf}
\put(32,28){$e$}
\put(40,49){$\gamma,\ Z$}
\put(40,09){$ \gamma, \ Z$}
\put(77,51){$ l$}
\put(77,35){${\bar l}$}
\put(77,21){${ b}$}
\put(77, 5){${\bar b}$}
\end{Feynman}
\end{center}
}\end{minipage}
\begin{minipage}[tbh]{7.8cm} {
\begin{center}
\begin{Feynman}{75,60}{0,0}{0.8}
%
\put(15,7.5){\fermiondllhalf}
\put(15,52.5){\fermiondrrhalf}
\put(30,15){\fermiondown}
\put(30,45){\photonright}
\put(30,15){\photonright}
\put(30,45){\circle*{1.5}}
\put(30,15){\circle*{1.5}}
\put(60,15){\circle*{1.5}}
\put(60,45){\circle*{1.5}}
\put(75,7.5){\fermionullhalf}
\put(75,22.5){\fermionurrhalf}
\put(75,37.5){\fermionullhalf}
\put(75,52.5){\fermionurrhalf}
\put(32,28){$e$}
\put(40,49){$\gamma,\ Z$}
\put(40,09){$\gamma,\ Z$}
\put(77,51){$ b$}
\put(77,35){${\bar b}$}
\put(77,21){${ l}$}
\put(77, 5){${\bar l}$}
\end{Feynman}
\end{center}
}\end{minipage}
\vspace*{.2cm}
\caption
{\it
\label{crabsd1}
The {\tt crab} diagrams.
}
\end{figure}

\begin{figure}[bhtp]
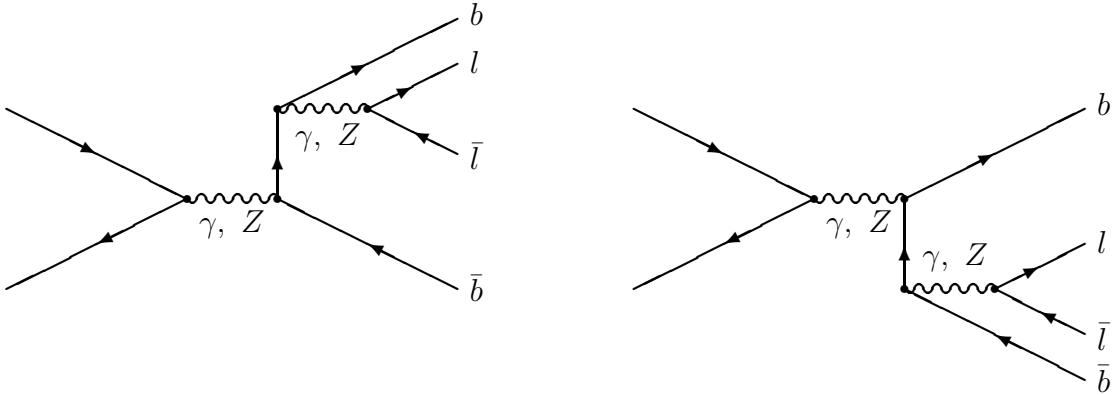

\begin{minipage}[tbh]{7.8cm}{
\begin{center}
\begin{Feynman}{75,60}{-5.0,0}{0.8}
%
\put(-10,45){\fermiondrr}
\put(-10,15){\fermiondll}
\put(20,30){\photonrighthalf}
\put(65,15){\fermionull}
\put(20,30){\circle*{1.5}}
\put(35,30){\circle*{1.5}}
\put(35,45){\circle*{1.5}}
\put(50,45){\circle*{1.5}}
\put(35,30){\fermionuphalf}
\put(65,60){\fermionurr}
\put(35,45){\photonrighthalf}
\put(65,52.5){\fermionurrhalf}
\put(65,37.5){\fermionullhalf}
\put(22,24.5){$\gamma,\ Z$}
\put(38,39){$\gamma,\ Z$}
\put(67,59){$ b$}
\put(67,51){$ l$}
\put(67,35){${\bar l}$}
\put(67,13){${\bar b}$}
\end{Feynman}
\end{center}
}\end{minipage}
\begin{minipage}[tbh]{7.8cm}{
\begin{center}
\begin{Feynman}{75,60}{0,0}{0.8}
%
\put(0,45){\fermiondrr}
\put(0,15){\fermiondll}
\put(30,30){\photonrighthalf}
\put(30,30){\circle*{1.5}}
\put(45,30){\circle*{1.5}}
\put(45,15){\circle*{1.5}}
\put(60,15){\circle*{1.5}}
\put(45,15){\fermionuphalf}
\put(75,00){\fermionull}
\put(75,45){\fermionurr}
\put(45,15){\photonrighthalf}
\put(75,22.5){\fermionurrhalf}
\put(75,07.5){\fermionullhalf}
\put(32,24.5){$\gamma,\ Z$}
\put(48,19){$\gamma,\ Z$}
\put(77,44){$ b$}
\put(77,21){$ l$}
\put(77, 5){${\bar l}$}
\put(77,-2){${\bar b}$}
\end{Feynman}
\end{center}
}\end{minipage}
\vspace*{.2cm}
\caption{\it
\label{ldeersd1}
The {\tt b-deer} diagrams. The {\tt l-deers} may be obtained by
interchanging the leptons with the quarks.
}
\end{figure}

\noindent
In~(\ref{domega}), the rotation angle around the beam axis has been
integrated over already.
Variables $k_1$ and $k_2$ are the four-momenta of
electron and positron and $p_1,p_2,p_3,p_4$ are those of the
final state particles $f_1,\bar{f_1}, f_2,\bar{f_2}$ with $p_i^2=m_i^2$.
All fermion masses are neglected compared to the invariants $s, s_1$, and
$s_2$~
\footnote{ 
An exception arises if there
are photon propagators $1/s_i$ involved and if no cut is applied.
In this case,
the finite masses of the fermions yield
non-logarithmic contributions of order ${\cal O}(1)$.
These may be taken
into account by replacing in~(\ref{domega}) the
$\sqrt{1-4m_f^2/s_i}$
by
$\sqrt{1-4m_f^2/s_i} \, (1+2m_f^2/s_{i})$ (instead of neglecting $m_f^2/s_i$).
}:
\begin{eqnarray}
s=(k_1+k_2)^2,\ \
s_1=(p_1+p_2)^2,\ \
 s_2=(p_3+p_4)^2.
\end{eqnarray}
The angle $\theta$ is located between the vectors ($\vec{p}_1+\vec{p}_2$) and
$\vec{k}_1$.
The spherical angle of $\vec{p_1}$
($\vec{p_3}$)
in the rest frame of the compound [$f_1{\bar f}_1$]
([$f_2{\bar f}_2$]) is
$\Omega_1\ (\Omega_2)$:
$d \Omega_i = d \cos\theta_i d \phi_i$.
The kinematical ranges of the integration variables are:
\begin{eqnarray}
(m_1+m_2)^2 \le s_1 \le (\sqrt{s}-m_3-m_4)^2&,&\ \ \
(m_3+m_4)^2 \le s_2    \le (\sqrt{s}-\sqrt{s_1})^2 ,
\nll
-1 \le \cos\theta,\ \cos\theta_1,\ \cos\theta_2 \le 1 &,&\ \ \
0 \le \phi_2,\phi_1 \le 2\pi.
\end{eqnarray}
%
We are interested in analytical formulae for distributions in invariant masses
of fermion pairs.
Thus, we have to integrate analytically over the five angular
variables in (\ref{domega}).
The matrix elements squared have been determined
by two independent calculations.
One of them used the symbolic manipulation program {\tt FORM}~\cite{form}.
The other one determined the
squared matrix elements with {\tt CompHEP}~\cite{comphep}
and continued then with the angular integrations in a different
{\tt FORM} program.

The total cross section
may be written as a sum of six contributions:
\ba
\sigma(s) =
\int_{{\bar s}_1}^s ds_1
\int_{{\bar s}_2}^{(\sqrt{s}-\sqrt{s_1})^2} ds_2
\, 
\sum_{k=1}^6
\frac{d^2\sigma_k(s,s_1,s_2)}{ds_1ds_2}.
\label{tsig}
\ea
In~(\ref{tsig}) we allow for lower cuts on the invariant masses.
Index $k=1$ corresponds to the square of the sum of the {\tt crabs},
and $k=2,3$
to that of the {\tt b-deers} and {\tt l-deers}.
Further, $k=4,5$ and 6 correspond to the interferences between
{\tt l-deers} and {\tt b-deers}, between
{\tt{crabs}} and {\tt b-deers},
 and between
{\tt{crabs}} and {\tt l-deers}, respectively.
%

\section{The cross section}
All the diagrams which mediate process~(\ref{lb}) have a common topology.
They contain three fermion antifermion pairs connected by neutral gauge
bosons.
Two of the fermion pairs couple to a gauge boson while the third one couples to
two gauge bosons with a fermion propagator between them.
In this sense, there is only one generic topology, the {\tt deer} topology,
and the {\tt crab} diagrams may be considered as the {\tt e-deers} in a crossed
channel.

Our choice of coordinates takes the above symmetry into account.
All the 300
interferences between the 24 diagrams may be reduced
to two generic ones.
This may be achieved by grouping the diagrams in the calculation as is
indicated in the figures, i.e. in sums of {\tt deers} of the same type.
A tremendous reduction of terms arises during the calculation
leading to an extremely compact result.

The first generic interference
is known from off shell $Z$ boson pair production~\cite{wwteup}.
It is the square of the {\tt crab} diagrams or generally of same-type {\tt
deers}. The three cases differ from each other by permutations of the invariant
masses and couplings.
The second generic kinematic function
is due to the interference of
different
types of {\tt deers}.
Again, three such cases are related by permutations.
Hence, the complete
cross section can be written as a sum of six basic interferences between
the three basic sets of diagrams.


The cross section from the two ${\tt crab}$ diagrams alone reads as follows:
\ba
\frac{d^2\sigma_1(s;s_1,s_2)}{ds_1ds_2} &=&
 \frac{\sqrt{\lambda}}{\pi s^2}
 C_{422}(e,s;l,s_1;b,s_2) {\cal G}_{422}(s;s_1,s_2)~.
\label{twocrabsg}
\ea
This expression factorizes into a
function of couplings and a kinematical
${\cal G}$-function
and is symmetric in $s_1$ and $s_2$;
the latter depends exclusively on the virtualities\footnote
{
This function may be found in the article~\cite{bfk} on electromagnetic
pair production and also in~\cite{cl} where extra neutral
 gauge boson production is
studied.
}:
\bq
{\cal G}_{422}(s;s_1,s_2) =
 \frac{s^2+(s_1+s_2)^2}{s-s_1-s_2} {\cal L}(s;s_1,s_2) -2,
\label{muta1}
\eq
with
\bq
{\cal L}(s;s_1,s_2) =
\frac{1}{\sqrt{\lambda}} \, \ln \frac{s-s_1-s_2+\sqrt{\lambda}}
                                     {s-s_1-s_2-\sqrt{\lambda}}.
\label{L4}
\eq

The couplings and gauge boson propagators
are contained in the following $C$-factor:
\ba
\label{c422}
C_{422}(e,s;f_1,s_1;f_2,s_2) &=& \frac{2 s_1 s_2}{(6\pi^2)^2} \;
  \Re e   \sum_{V_i,V_j,V_k,V_l=\gamma,Z}
\frac{1}{D_{V_i}(s_1)}      \frac{1}{D_{V_j}(s_2)}
\frac{1}{D_{V_k}^*(s_1)}    \frac{1}{D_{V_l}^*(s_2)}
\nll & &
\times~
\left[ L(e,V_i)L(e,V_k)L(e,V_j)L(e,V_l)
+R(e,V_i)R(e,V_k)R(e,V_j)R(e,V_l)\right]
\nll & &
\times~
\left[ L(f_1,V_i)L(f_1,V_k)+R(f_1,V_i)R(f_1,V_k)\right] N_c(f_1)
\nll & &
\times~
\left[ L(f_2,V_j)L(f_2,V_l)+R(f_2,V_j)R(f_2,V_l)\right] N_c(f_2).
\ea
The conventions for the left- and right-handed couplings
between vector bosons and fermion $f$ are:
\ba
\label{couplings}
 L(f,\gamma) &=& R(f,\gamma) =  \frac{e Q_f}{2},
\\
 L(f,Z) &=& \frac{e}{4 s_W c_W}\left(2 I_3^f - 2 Q_fs^2_W \right),\ \ \
 R(f,Z) = \frac{e}{4 s_W c_W}\left(-2 Q_f s^2_W \right).
\ea
Here, we use
$e=\sqrt{4\pi\alpha}$,
$Q_e=-1$, $I_3^e=-\frac{1}{2}$, and $s^2_W = 1-M_W^2/M_Z^2$.
Further, we choose $\alpha / \pi$ $=$ $ \sqrt{2} M_W^2 s_W^2 G_{\mu} / \pi^2$.
The numerical input for the figures is $G_{\mu} = 1.16639 \times
10^{-5}$, $M_W = 80.220$ GeV, $M_Z = 91.173$ GeV,
$\Gamma_Z=2.497$ GeV.
The colour factor $N_c(f)$ is equal to unity for leptons and three for
quarks.
The propagators are
\bq
D_V(s)=s-M_V^2+i\sqrt{s} \, \Gamma_V(s).
\label{propagator}
\eq
For the photon, it is
$M_{\gamma}=\Gamma_{\gamma}=0$.
The $Z$ width function depends on the decay channels which are open at a given
energy; at the $Z$ peak and
above, a good approximation is
$\Gamma_Z(s)=\sqrt{s}\, \Gamma_Z / M_Z$.
The
$D^*$ is the
complex conjugate of the propagator $D$.
The resonating behaviour of the cross section depends on the arguments
of the function $C_{422}$.
With e.g. all $V_i=Z$ in~(\ref{c422}),
one selects the off shell $Z$ pair production diagrams squared
(see figure~1), whose contribution is
proportional to two Breit Wigner resonance factors,
\ba
C_{422} = \ldots +
2
\left[ L^4(e,Z) + R^4(e,Z)\right]
\, \,    \frac{1}{\pi} \, \frac{\sqrt{s_1}\,  \Gamma_Z(f_1)}{|D_Z(s_1)|^2}
\, \,    \frac{1}{\pi} \, \frac{\sqrt{s_2}\,  \Gamma_Z(f_2)}{|D_Z(s_2)|^2},
\label{breit}
\ea
with
\ba
\lim_{\Gamma \, \rightarrow 0}\,
\frac{1}{\pi} \frac{\sqrt{s}\,  \Gamma}{|D(s)|^2} = \delta(s-M^2).
\label{delta}
\ea

The cross section contributions $\sigma_2$ from the square of the
{\tt b-deers} and $\sigma_3$
of the {\tt l-deers} are:
\ba
\frac{d^2\sigma_2(s;s_1,s_2)}{ds_1ds_2} &=&
 \frac{\sqrt{\lambda}}{\pi s^2}
C_{422}(b,s_2;e,s;l,s_1) {\cal G}_{422}(s_2;s,s_1),
\label{2bbdeersg}
\\ 
\frac{d^2\sigma_3(s;s_1,s_2)}{ds_1ds_2} &=&
 \frac{\sqrt{\lambda}}{\pi s^2}
C_{422}(l,s_1;b,s_2;e,s) {\cal G}_{422}(s_1;s_2,s).
\label{2lldeersg}
\ea
In these contributions the potential resonance behaviour is in the variables
$s$ and $s_1$ {\em or} $s_2$, respectively.

Besides the above moduli squares, there are interferences among the three
different groups of diagrams.
The interference between {\tt l-deers} and {\tt b-deers} is again symmetric in
the last two arguments:
\ba
\frac{d^2\sigma_4(s;s_1,s_2)}{ds_1ds_2} &=&
 \frac{\sqrt{\lambda}}{\pi s^2}
C_{233}(e,s;l,s_1;b,s_2) {\cal G}_{233}(s;s_1,s_2),
\label{intdeersg}
\ea
with the kinematical function ${\cal G}_{233}(s;s_1,s_2)$,
\ba
 {\cal G}_{233}(s;s_1,s_2) &=&
 \frac{3}{\lambda^2}\Biggl\{
 {\cal L}(s_2;s,s_1) {\cal L}(s_1;s_2,s)
          4s\left[ss_1(s-s_1)^2+ss_2(s-s_2)^2+s_1s_2(s_1-s_2)^2\right]
\nll
& & +~(s+s_1+s_2)\Biggl[
            {\cal L}(s_2;s,s_1)
          2s\left[(s-s_2)^2+s_1(s-2s_1+s_2)\right] \nll
& & \hspace{2.5cm} +~ {\cal L}(s_1;s_2,s)
          2s\left[(s-s_1)^2+s_2(s+s_1-2s_2)\right] \nll
& & \hspace{4.9cm} +~ 5s^2-4s(s_1+s_2)-(s_1-s_2)^2 \Biggr] \Biggr\}.
\label{muta2}
\ea
For $\lambda \rightarrow 0$ it remains finite,
${\cal G}_{233}(s;s_1,s_2)$
$\rightarrow$
$-3$
$(3s - s_1 - s_2)/$
$[(s+s_1-s_2)(s+s_2-s_1)]$.
The $C$ factor with couplings and propagators is:
\ba
\label{c233}
C_{233}(e,s;f_1,s_1;f_2,s_2) &=&
 \frac{2 s s_1 s_2}{(6\pi^2)^2}
\, \Re e   \sum_{V_i,V_j,V_k,V_l=\gamma,Z}
\frac{1}{D_{V_i}(s)}\frac{1}{D_{V_j}(s_2)}
\frac{1}{D_{V_k}^*(s)}\frac{1}{D_{V_l}^*(s_1)}
\nll & &
\times~
\left[ L(e,V_i)L(e,V_k)+R(e,V_i)R(e,V_k)\right]
\nll & &
\times~
\left[ L(f_1,V_i)L(f_1,V_j)L(f_1,V_l)-R(f_1,V_i)R(f_1,V_j)R(f_1,V_l)\right]
N_c(f_1)
\nll & &
\times ~
\left[ L(f_2,V_j)L(f_2,V_k)L(f_2,V_l)-R(f_2,V_j)R(f_2,V_k)R(f_2,V_l)
\right] N_c(f_2).
\nll
\ea
Both final fermion traces must couple to an axial current. Otherwise
the contribution vanishes due to the Furry theorem.

The remaining two interferences $\sigma_5$ among {\tt crabs} and {\tt b-deers}
and $\sigma_6$ among {\tt crabs} and {\tt l-deers} are:
\ba
\frac{d^2\sigma_5(s;s_1,s_2)}{ds_1ds_2} &=&
 \frac{\sqrt{\lambda}}{\pi s^2}
C_{233}(l,s_1;b,s_2;e,s) {\cal G}_{233}(s_1;s_2,s),
\label{int2ld2cg}
\ea
\ba
\frac{d^2\sigma_6(s;s_1,s_2)}{ds_1ds_2} &=&
 \frac{\sqrt{\lambda}}{\pi s^2}
C_{233}(b,s_2;e,s;l,s_1) {\cal G}_{233}(s_2;s,s_1).
\label{int2bd2cg}
\ea

The interferences of different-type {\tt deers} may become resonating
only
in the first argument of the function $C_{233}(f_1,s_1;f_2,s_2;f_3,s_3)$.

\bigskip

At the end of this section, we would like to comment on processes with four
quarks in the final state.
In this case, besides photons and $Z$ bosons also gluons may be exchanged in
the {\tt deer} diagrams while
the {\tt crab} contribution $\sigma_1$ remains unchanged.
The gluonic contributions to $\sigma_5$ and $\sigma_6$
vanish due to the colour trace, but
$\sigma_2$, $\sigma_3$, and $\sigma_4$ get additional contributions.
The sum over gauge bosons in the
$C$-functions  in (\ref{2bbdeersg})--(\ref{intdeersg})
extends now also over gluons.
The gluon couplings to quarks are
\bq
\label{couplingsqcd}
 L(q,g) = R(q,g) =
\frac{1}{2} \sqrt{4\pi\alpha_s},
\eq
and the gluon propagator looks like the photon propagator.
Further, the colour factor $N_c(f_1) N_c(f_2)$ has to be
replaced by a factor of $(N_c^2-1)/4=2$
 for all interferences with {\it two} gluons and
by a zero if only {\it one} gauge boson is a gluon.

If one observes two hadron jets without flavour tagging, then there are
also
contributions
from diagrams with a three gluon vertex which have to be added incoherently.
If the quark types may be tagged, such contributions are of higher order
and may be neglected.

\section{Results}
%
Numerical comparisons
have been performed
for $e^+e^-\rightarrow\mu^+\mu^-b\bar{b}$
and  $e^+e^-\rightarrow\nu_\mu\bar{\nu}_\mu b\bar{b}$ at LEP~2 energies
with a Monte Carlo approach~\cite{pittau}; see also~\cite{bglasg}.
The agreement is within 0.2\% and is limited by the
numerical accuracy of the Monte Carlo program.

There are two different energy regions of physical interest.
At LEP~1, among others the cases~(i) and~(ii--3) (see Introduction) are
observed~\cite{lep1zz} and
case (ii--1) is searched for as a Higgs boson signal~\cite{lep1zh}\footnote{
We did not perform a dedicated comparison with the experimental results
at LEP~1 \cite{lep1zz}
since
the applied cuts are different from ours;
our predictions agree roughly with them.}.
While, at higher energies the most interesting channel is
{}~(ii--1).

In figure~\ref{f3}, the cross sections
of cases~(i),~(ii--1), and~(ii--3) are
shown at LEP~1
energies.
We assume cuts being applied to the final state fermion pairs.
For illustrational purposes, cross section (i) is shown also without
cut.
The cross section is sensitive to this due to the diagrams with photon
exchange
which are extremely enhanced when their virtualities approach to
$4m_f^2$.
The cross sections are peaking at $\sqrt{s} = M_Z$,
if the non-resonating
gauge
boson is a photon with a small
virtuality.
In~\cite{kleiss2}, partial widths of the $Z$ boson into $4f$ final states are
systematically studied.
Having in mind that the production cross sections are proportional to them
at the $Z$ peak, one may compare corresponding ratios with each other.
We have done this in the massless limit
for the cases
[$\mu \bar \mu, \tau \bar \tau$],
[$\mu \bar \mu, b \bar b$],
[$\mu \bar \mu, (u \bar u + d \bar d + s \bar s + c \bar c + b \bar b)$] and
found
agreement within 1~\% which is the accuracy of the numbers quoted
in~\cite{kleiss2}.

\begin{figure}[thbp]
\begin{center}
  \vspace{-0.5cm}
  \hspace{-2.3cm}
  \mbox{
  \epsfysize=9cm
  \epsffile[0 0 500 500]{leppub.ps}
  }
\end{center}
 \vspace{-1.2cm}
\caption{\it
\label{f3}
Total cross sections $\sigma(e^+e^- \rightarrow 4f)$ at LEP~1
energies.
The following cuts are applied:
$E_{q \bar q} ,
E_{\mu \bar \mu} > 2$ {\rm GeV},
$E_{\tau \bar \tau} > 2 m_{\tau}$,
$E_{c \bar c} > 5$ {\rm GeV},
$E_{b \bar b} > 20$ {\rm GeV}.
}
\end{figure}
\begin{figure}[bhtp]
\begin{center}
\vspace{-0.5cm}
\hspace{-2.3cm}
\mbox{
\epsfysize=9cm
\epsffile[0 0 500 500]{nlcpub.ps}
}
\end{center}
 \vspace{-1.2cm}
\caption{
\label{f4}
\it
Total cross sections $\sigma(e^+e^- \rightarrow 4f)$ at LEP~2 and NLC
energies.
The following cuts are applied:
$E_{\mu \bar \mu} > 2$ {\rm GeV},
$E_{\tau \bar \tau} > 2 m_{\tau}$,
$E_{c \bar c} > 5$ {\rm GeV},
$E_{b \bar b} > 20$ {\rm GeV}.
}
\end{figure}

Figure~\ref{f4} shows
final states~(i),~(ii--1),
and~(iii--1)   in
                 the             energy  region
$\sqrt{s} \sim 100 - 500$ GeV.
Again, in one case the influence of the cuts on the invariant masses
is exhibited.
As to be expected, there are pronounced
thresholds at the onset of on shell $Z$ pair
production.  At higher energies,
the cross sections fall monotonically.
In the figure, we show also a  four quark final state.
The strong coupling constant $\alpha_s$ is a function of the
virtualities $s_1$ or $s_2$ (and in some interferences of both)
 and may not be reasonably chosen here.
For the illustrational purposes, we fixed it somehow arbitrarily at
$\alpha_s=0.2$; this corresponds to smaller values of the gluon
virtualities which give the dominant contributions.

We also should mention that there are substantial radiative corrections
to $4f$ production.
The bulk of them is known to arise from initial state photonic bremsstrahlung.
It may be easily taken into account with a convolution formula which is
well known from the $Z$ line shape~\cite{pittau,wwteup}.
These corrections amount to -- ${\cal O}(15\%)$ around the peaks and are
smaller
elsewhere, becoming positive (and cut dependent) in the radiative tail
regions.
Since all these effects are well known we do not go into numerical details
on
them.
Our Fortran program {\tt 4fAN} may take them into account.

\bigskip

{\em To summarize},
we performed the first complete semi-analytical calculation
of neutral current four fermion production.
We obtained, for the simplest topology,
                                               extremely compact
analytical results for the two-dimensional invariant mass distributions.
The remaining two integrations are fast and numerically stable.
Polarized beams and the inclusion of heavy neutral gauge bosons
are also described  by our formulae.

\section*{Acknowledgement}
We would like to thank F.~Berends for discussions, hints, and
a numerical comparison.


\end{document}